# SK-Unet: an Improved U-net Model with Selective Kernel for the Segmentation of Multi-sequence Cardiac MR

Xiyue Wang[1#], Sen Yang[1,4#], Mingxuan Tang[2], Yunpeng Wei[3], Ling He[1(✉)], Jing Zhang[1(✉)], Xiao Han[4]

[1]College of Electrical Engineering, Sichuan University, Chengdu, China
[2] The School of Computer Science, Chengdu University of Information Technology, Chengdu, China
[3]Chinese Academy of Medical Sciences and Peking Union Medical College, Beijing, China
[4]Tencent AI Lab, Shenzhen, China

**Abstract.** In the clinical environment, myocardial infarction (MI) as one common cardiovascular disease is mainly evaluated based on the late gadolinium enhancement (LGE) cardiac magnetic resonance images (CMRIs). The automatic segmentations of left ventricle (LV), right ventricle (RV), and left ventricular myocardium (LVM) in the LGE CMRIs are desired for the aided diagnosis in clinic. To accomplish this segmentation task, this paper proposes a modified U-net architecture by combining multi-sequence CMRIs, including the cine, LGE, and T2-weighted CMRIs. The cine and T2-weighted CMRIs are used to assist the segmentation in the LGE CMRIs. In this segmentation network, the squeeze-and-excitation residual (SE-Res) and selective kernel (SK) modules are inserted in the down-sampling and up-sampling stages, respectively. The SK module makes the obtained feature maps more informative in both spatial and channel-wise space, and attains more precise segmentation result. The utilized dataset is from the MICCAI challenge (MS-CMRSeg 2019), which is acquired from 45 patients including three CMR sequences. The cine and T2-weighted CMRIs acquired from 35 patients and the LGE CMRIs acquired from 5 patients are labeled. Our method achieves the mean dice score of 0.922 (LV), 0.827 (LVM), and 0.874 (RV) in the LGE CMRIs.

## 1　Introduction

Cardiac magnetic resonance images (CMRIs) with the capacity of discriminating various types of tissues are primarily used in the diagnosis and treatment of cardiovascular diseases, such as the myocardial infarction (MI). Late gadolinium enhance-ment (LGE) cardiac magnetic resonance (CMR) sequence has the capacity to visualize the infarcted myocardium that has the enhanced brightness compared with the healthy myocardium [1, 2]. Thus, accurate delineation of the left ventricle (LV), right ventricle (RV), and left ventricular myocardium (LVM) in the LGE CMRIs is important in the clinical diagnosis. The manual delineation by a clinical radiologist for these three parts is time-consuming, tedious, and rater-dependent [1, 3]. Hence, an automatic segmentation method for this task is desirable in clinical setups.

---

✉　Corresponding authors
\#　Co-first authors

Currently, a majority of studies for the automatic cardiac segmentation are based on the cine CMR sequence [3-8], since the cine CMR sequence has the ability to capture the cardiac motions during the whole cardiac cycle and can present clear boundary [1]. The LGE CMR sequence enhances the representation of the infarcted myocardium and is routinely used in the clinical diagnosis of MI. However, there is few research on automatic cardiac segmentation directly in the LGE CMR sequence, since the CMRIs with poor image quality have heterogenetic intensity distribution. The current LGE CMR based cardiac segmentation research mainly targets to delineate the contour of LVM [1, 9-14]. They accomplish the segmentation task utilizing a semi-automated approach [12] or combining the prior segmentation contour in the corresponding cine CMR sequence from the same patient with the same phase [9-11, 13, 14]. The prior knowledge based methods usually require the assistance of various image registration algorithms. The result of image registration may produce errors due to the varied slice thickness and spatial resolution in different patients.

To the best of our knowledge, there is almost no research focusing on the simultaneously automatic segmentation of the LV, RV, and LVM in the LGE CMRIs. Manual delineation in the LGE CMRIs is particularly arduous. This paper proposes an automatic segmentation algorithm for these three parts in the LGE CMRIs. The utilized dataset is from the MICCAI challenge (MS-CMRSeg 2019), which is collected from 45 patients. Only 5 of 45 patients have their LGE CMRIs labeled, and 35 of 45 patients have their cine and T2-weighted CMRIs labelled. In this paper, the LGE CMRIs with a small amount of manual delineations are segmented by combing anther two CMR sequences (cine and T2-weighted CMRIs) acquired from the same patient. This paper achieves this segmentation based on a modified U-net architecture. The squeeze-and-excitation residual (SE-Res) [15] and selective kernel (SK) modules [16] are respectively inserted in the down-sampling and up-sampling stages of the conventional U-net architecture. The SE-Res module considers more channel dependencies and lacks the spatial information of feature maps. The spatial information is important for the pixel-level localization in the image segmentation task. The SK module is utilized to relieve this problem, which adaptively adjusts the size of local respective field in the convolutional operation to collect multi-scale spatial information [16]. The proposed SK-Unet framework has achieved robust segmentation performance in the LGE CMR sequences.

## 2   Methodology

The modified architecture used for this cardiac segmentation task is based on the classical U-net architecture. The proposed LV, RV, and LVM segmentation algorithm includes three parts: image preprocessing, SK-Unet model based image segmentation, and image postprocessing.

### 2.1   Image Preprocessing

In order to remove the influence of the surrounding organ of heart in the CMRIs, region of interest (ROI) extraction is a crucial step in the prepossessing stage. The distribution range of intensity, image contrast, and image size are different in these three CMR sequences. It is difficult to develop a robust ROI detection method. Since the three CMR sequences of each patient are acquired in same session and with the same cardiac phase, the anatomical structure is consistent in the three CMR sequences. Thus, this paper performs a statistical work to roughly locate the position of the heart.

The second step in the preprocessing process is to normalize the input images as the distribution of zero mean and variance of 1. Then, due to the limited training images, the data augmentation is applied to create an expanded dataset from the original dataset. The adopted data augmentation methods include image transpose, flipping, cropping, and rotation. Finally, to fully utilize the dependences between slices, the neighbored three slices are stacked as the new three-channel image that has the same mechanism as the RGB channel in the color image.

### 2.2   SK-Unet Based Image Segmentation Model

The proposed SK-Unet based CMR segmentation model consists of the encoding and decoding stages. The skip connections exist between encoder and decoder blocks with the same image spatial resolution. Fig. 1 illustrates the overall structure of the proposed SK-Unet based CMR segmentation model.

As shown in Fig. 1(a), the input is the neighbored three slices in one CMRI, and the output is the probability that each pixel in the CMRI is classified as the background, LV, RV, and LVM. The left part performs the encoding operation with pooling layer, convolution layers, and SE-Res module, and the decoding operation with up-sampling unit, convolution layers, and SK module is performed in the right part. The horizontal connections mean that the extracted features in the encoding stage are forwarded to the corresponding decoding stage.

The pooling layers reduce the spatial resolution of feature maps to attain high-level feature representation. As shown in Fig. 1(a), the stride in pooling layers is adopted as 2. The up-sampling unit has the inverse operation as the pooling layer.

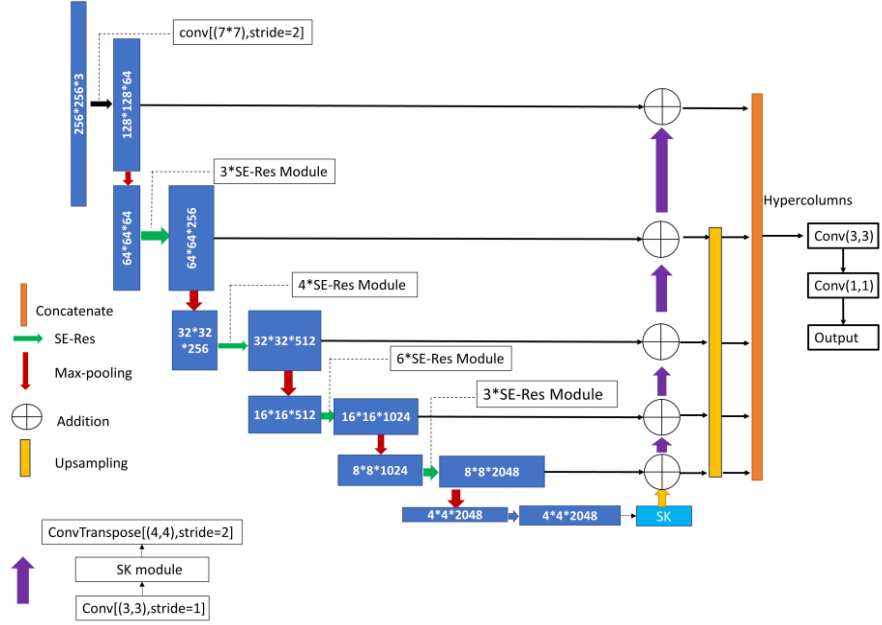

(a)

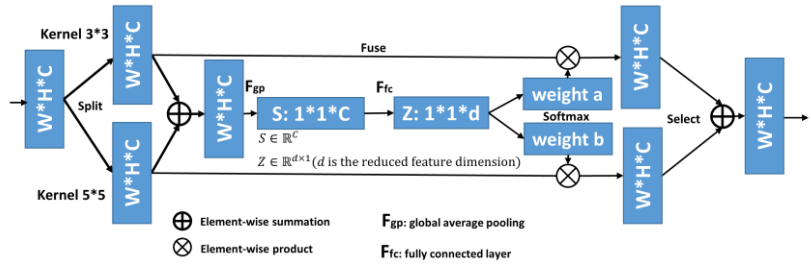

(b)

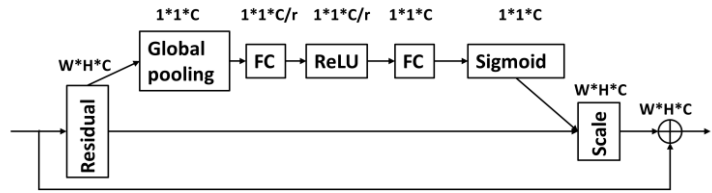

(c)

**Fig. 1.** The overall structure of the proposed SK-Unet based CMR segmentation model. The SK and SE-Res modules are shown in (b) and (c).

The convolution operation at each convolution layer applies filters to learn informative features by combining the spatial and channel-wise information within the local receptive fields. In order to attain global feature fusion, the SE-Res and SK modules are inserted in the conventional U-net architecture. The SE-Res module factors out the spatial correlations among features and captures the relationships among channels, which is effective in the classification task. In this CMR image segmentation task, the spatial information with the representation of texture, boundary, and gray-level, is also important. The SK module can adaptively adjust the size of local respective field in the procedure of the CNN operation, which helps capture multi-scale spatial correlations among features. Thus, the SK module makes the obtained feature maps more informative in both spatial and channel-wise space. The proposed architecture has robustness for the classification between the target region and background, and for the precise localization of the target region in the CMR image segmentation task.

### 2.3 Image Postprocessing

The image postprocessing process aims to refine the result of the cardiac segmentation. First, the hole filling technique is applied to attain more complete segmentation. Then, a connected component analysis for all the obtained segmentation is performed. The largest connected component of all slices in each patient is found, which is set as the constraint for modifying the segmentation of the remaining slices in each patient. Segmentations that exceed the largest connected range will be removed.

## 3 Experimental and Results

### 3.1 Dataset: MS-CMRSeg 2019

Our algorithm is evaluated on the Multi-sequence Cardiac MR Segmentation Challenge MICCAI 2019 (MS-CMRSeg 2019) dataset [1, 2]. This dataset covers 45 patients with cardiomyopathy, and each patient has been scanned by the cine, T2-weighted, and LGE CMR sequences from the short-axis orientation. The cine CMRIs acquired with the balanced-steady state free precession (bSSPF) sequence covers the full ventricles, which are selected in the same cardiac phase as the following LGE and T2-weighted CMRIs. The LGE and T2-weighted CMRIs cover the main body of the ventricles, which are collected at the end-diastolic phase. Since the number of labelled samples is limited, the cine and the T2-weighted CMRIs collected from 35 patients, and the LGE CMRIs collected from 5 patients are used as the training data. The testing data adopts the LGE CMRIs collected from 40 patients.

### 3.2 The Overall Performance of the Proposed Approach

Four commonly used indicators in medical image segmentation are used to evaluate the performance of the model. These four indicators include dice score, Hausdorff distance, average surface distance, and Jaccard index, which are listed in Table 1.

**Table 1.** The mean and standard deviation of the dice score, Hausdorff distance, surface distance, and Jaccard index in the cardiac segmentation task.

|  | Mean ± standard deviation | | |
| --- | --- | --- | --- |
|  | LV blood cavity | LV myocardium | RV blood |
| Dice score (%) | 0.922 ± 0.036 | 0.827 ± 0.060 | 0.874 ± 0.058 |
| Jaccard index (%) | 0.857 ± 0.059 | 0.709 ± 0.084 | 0.781 ± 0.089 |
|  | LV endocardium | LV epicardium | RV endocardium |
| Hausdorff distance (mm) | 10.058 ± 3.820 | 11.426 ± 3.574 | 16.721 ± 7.509 |
| Surface distance (mm) | 1.589 ± 0.637 | 1.696 ± 0.585 | 2.208 ± 1.016 |

As shown in Table 1, the segmentation for the LV cavity reaches the highest performance in term of dice score, Jaccard index, Hausdorff distance, and surface distance. The segmentation of LV myocardium is relatively difficult since the existence of the invalid tissue. The invalid tissue, for instance the MI, has the same appearance as the blood pool, which results in the difficulty in the localization of the LV myocardium.

### 3.3 Comparison with Various Image Segmentation Architectures

In order to intuitively represent the segmentation performance of the proposed algorithm, the visual segmentation results for the LV, LVM, and RV are compared with various architectures including DenseUnet [17], Linknet [18], and U-net [19], which are represented in Fig. 2.

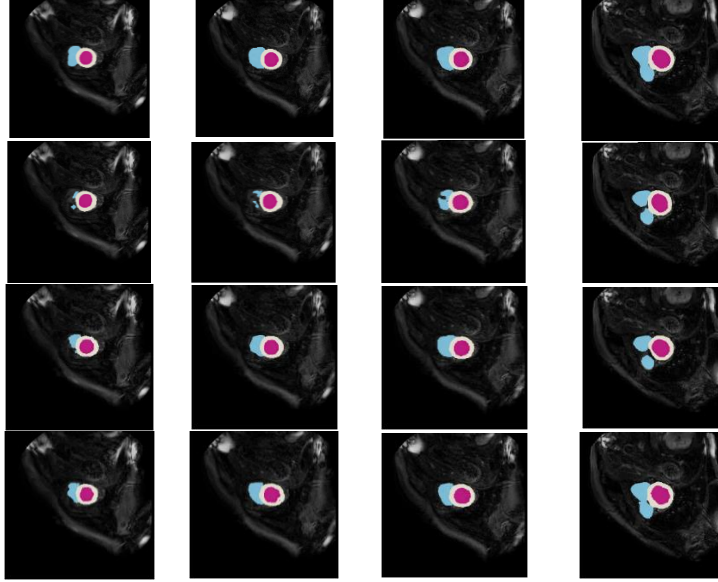

**Fig. 2.** The visual segmentation results for the LV, LVM, and RV of the same slices from one specific patient using various architectures. The results from the first to the last columns are obtained using SK-Unet, DenseUnet, Linknet, and U-net, respectively.

As shown in Fig. 2, the SK-Unet based algorithm achieves the best segmentation result. The Linknet and U-net based methods produce unsmooth segmentation in the boundary. The DenseUnet has difficulty in segmenting the RV part.

The SK-Unet module fully utilizes the information of channels in the feature maps through the inserted SE-Res module. Meanwhile, the adaptive adjust of the receptive field size helps capture multi-scale spatial information. The combination of the channel and spatial information could learn more inter-slices and intra-slice features, and then produce more precise segmentation result.

## 4      Conclusion

This paper proposes an approach for the multi-sequence ventricle and myocardium segmentation using deep learning technique. We employ a modified U-net architecture with SE-Res and SK model. The SK model with concurrent spatial and channel information is beneficial for the target localization and pixel-level based classification in this segmentation task. These results suggest that our approach has the potential to provide aided diagnoses for clinical cardiac surgeon.